\documentclass{aa}
\usepackage{graphicx}
\usepackage{txfonts}
\usepackage{xcolor}
\usepackage{float}
\usepackage{hyperref}
\usepackage{amsmath}
\usepackage{nccmath}
\usepackage{array}
\usepackage{fontawesome}

\newcommand{\github}{\href{https://github.com/CosmoStat/BlendHunter}{\faGithub}}

\begin{document} 

\title{Deep Transfer Learning for Blended Source Identification in Galaxy Survey Data}

\author{S. Farrens\inst{\ref{cosmostat}} \thanks{email: samuel.farrens@cea.fr}
		\and A. Lacan\inst{\ref{usp}} 
		\and A. Guinot\inst{\ref{cosmostat}}
		\and A. Z. Vitorelli\inst{\ref{cosmostat}}
		}

\institute{AIM, CEA, CNRS, Université Paris-Saclay, Universit\'{e} Paris Diderot, Sorbonne Paris Cité, F-91191 Gif-sur-Yvette, France\label{cosmostat} 
\and Universit\'{e} Paris-Saclay, CNRS, CEA, Astrophysique, Instrumentation et Modélisation de Paris-Saclay, 91191, Gif-sur-Yvette, France\label{usp} } 
	 
\date{}

\abstract{We present \textsc{BlendHunter}, a proof-of-concept for a deep transfer learning based approach for the automated and robust identification of blended sources in galaxy survey data. We take the VGG-16 network with pre-trained convolutional layers and train the fully connected layers on parametric models of COSMOS images. We test the efficacy of the transfer learning by taking the weights learned on the parametric models and using them to identify blends in more realistic CFIS-like images. We compare the performance of this method to \textsc{SEP} (a Python implementation of \textsc{SExtractor}) as function of noise level and the separation between sources. We find that \textsc{BlendHunter} outperforms \textsc{SEP} by $\sim 15\%$ in terms of classification accuracy for close blends ($<10$ pixel separation between sources) regardless of the noise level used for training. Additionally, the method provides consistent results to \textsc{SEP} for distant blends ($\geq10$ pixel separation between sources) provided the network is trained on data with a relatively close noise standard deviation to the target images. The code and data have been made publicly available to ensure the reproducibility of the results. 
\github}


\maketitle


\section{Introduction}

The blending of sources (\textit{i.e.} the apparent overlapping of extended objects in 2D images) has a significant impact on the measurement of the morphological and structural properties of galaxies, in particular for ground-based surveys. Nearby objects can easily be mistaken for a single source, which can lead to significant detection and/or measurement biases depending on the depth of the survey.

For weak gravitational lensing analysis, it is essential to understand how and to what degree blending impacts shear and photometric redshift measurements \citep{mandelbaum:18}. This problem becomes even more critical in the current epoch of high precision cosmology, where systematic effects need to be carefully accounted for. This is particularly important for reliable comparisons between late-time probes, like weak lensing, and early-time probes, such as the CMB.   

Blended sources make up a significant fraction of the observed sources in survey images: $>30\%$ for the Dark Energy Survey \citep[DES,][]{samuroff:18}, $>50\%$ (up to $>60\%$ and $>70\%$ for the Deep and UltraDeep layers, respectively) for Hyper Suprime-Cam \citep[HSC,][]{bosch:18} and $>60\%$ for Vera C. Rubin Observatory Legacy Survey of Space and Time \citep[LSST,][]{sanchez:21}. Simply removing all the blends that have been identified would significantly reduce the sample size and could potentially lead to entangled biases in the shear correlation function \citep{hartlap:11}. Additional biases can be introduced by unidentified blends coming from sources below the detection threshold \citep{hoekstra:17, martinet:2019}. It is therefore necessary to develop an appropriate procedure for dealing with blends in survey data.

The process of handling blended sources in astrophysical images can be broadly divided into several problems: a) the detection of objects in the image, b) the classification of those objects as either single sources or blended sources,  c) the segmentation of pixels from blended sources (often referred to as `deblending'), and d) the rejection of those objects that cannot be easily included in the scientific analysis. This paper focuses on the problem of classifying objects, already detected with standard source extraction software, as either blends or not. Unlike the problem of segmentation, which has been abundantly addressed in the literature \citep{joseph:16,melchior:18,reiman:19}, little effort has been made to find reliable and automated methods for identifying blended sources in survey data.

Traditional methods for identifying blended sources, such as \textsc{SExtractor} \citep{bertin:96}, rely on fixed thresholds to detect multiple peaks in the light intensity profiles of the objects. While this approach may work for a reasonable fraction of the sources, it lacks the flexibility to handle blends with a bigger discrepancy in the brightness and/or size of the individual objects. This means that a large number of blended sources, which may have a non-negligible impact on the scientific analysis, could be missed.

Machine learning techniques, in particular deep learning architectures, have been shown to be incredibly successful when applied to complicated classification problems (see \textit{e.g.} \citet{kotsiantis:07}, \citet{lecun:2015}, and \citet{srinivas:2016}). The effectiveness of these tools, however, can be difficult to gauge without reliable labelled data. In real astrophysical images it is not known a priori if  the underlying signal for a given detection comes from a single object or indeed from the combination of several. This necessitates the use of simulated data in order to generate a reliable training set, where the number and diversity of overlapping sources are perfectly known. This, however, introduces the possibility of over-fitting the network to properties specific to the simulation, \emph{e.g.} overly simplified galaxy models. This in turn can cause the network to be very sensitive to artefacts and more complex structures seen in real observations. It is extremely challenging and time consuming to attempt to develop simulated galaxy images that contain all of the properties expected in real data. Therefore, the application of networks well suited to transfer learning  presents an interesting approach to mitigate this problem.

Transfer learning, in particular deep transfer learning, is a machine learning approach whereby network weights obtained by training on a given data set are applied to another distinct but similar data set. This can help prevent over-fitting to the training data as well as significantly reducing the time required to apply the method to new data sets. Deep transfer learning has been applied to a variety of astrophysical applications in recent years including: the classification of compact star clusters \citep{wei:2020}, the separation of low surface brightness galaxies from artefacts \citep{tanoglidis:2020}, the classification of planetary nebulae \citep{awang:2020} and the deblending of galaxy images \citep{arcelin:2021}.

This work introduces a proof-of-concept deep transfer learning approach for the automated and robust identification of blended sources in single-band galaxy survey data, hereafter referred to as \textsc{BlendHunter} (BH)\footnote{In the spirit of reproducible research, all code and data needed to reproduce the results in this paper have been made publicly available on GitHub (\url{https://github.com/CosmoStat/BlendHunter}) without any restrictions.}. This method incorporates a Convolutional Neural Network (CNN) trained on a large database of natural images and a fully connected layer for classification. Simple parametric models are used to train the fully connected layers and the learned weights are in turn used to identify blended sources in more realistic images. 

This paper is organised as follows. The following section introduces the properties of the training and testing data. Sect.~\ref{sec:framework} presents the transfer learning approach used and how it was trained to identify blended galaxy images. Sect.~\ref{sec:comparison} provides results on how this approach compares to the state of the art. Finally, conclusions are presented in Sect.~\ref{sec:conclusions}.


\section{Data}
\label{sec:data}

Supervised machine learning approaches, such as that presented in this work, require accurately labelled and representative training data. These labels often correspond to properties that cannot be directly and/or reliably measured from observed data. Simulations, on the other hand, provide a controlled environment where these properties are known a priori.

To train our network we opted to produce a set of simulated single-band (\emph{i.e.} monochromatic) galaxy images from a parametric model. This allowed us to control the fraction of blended images in the whole sample and the separation of sources in the blended images. The choice of single-band images was made to restrict the learning to pixel features (\emph{i.e.} no colour information), which could potentially be of interest for the $r$-band of the Canada-France Imaging Survey (CFIS)\footnote{\url{https://www.cfht.hawaii.edu/Science/CFIS/}}, part of the Ultraviolet Near-Infrared Optical Northern Survey (UNIONS), or eventually the Euclid visible band \citep{cropper:12}. We additionally prepared a sample of realistic CFIS-like galaxy images to test the efficacy of transferring the learned weights to a similar but unseen data set.

This section details how the training and testing data sets were generated.

\subsection{Blend definition}
\label{sec:blend_def}

In order to label our training data, we first needed to make a clear definition of a blend. Typically this would be some measure of amount of overlap between the light profiles of the individual sources that constitute the blend. For the purposes of this work, we assume a simple scenario in which isolated (\emph{i.e.} un-blended) sources are composed of a single galaxy centred within a postage stamp of size $51\times 51$ pixels ($9.5\mathrm{arcsec}\times9.5\mathrm{arcsec}$). We then define a blend as a postage stamp (of the same size) containing two galaxies, one at the centre and  a secondary source at a random position. This definition is fairly agnostic in that we do not require any specific overlap between the light profiles. This choice was made in order to gauge classification accuracy as a function of distance between the two sources. 

Given the size of the postage stamps, we expect that standard source extraction software, such as \textsc{SExtractor}, will be easily able to identify and separate sources near the borders but may struggle as the sources get closer together. To better highlight this point we additionally separate blends into two categories: `close blends' and `distant blends'. We define close blends as postage stamps in which the two sources are separated by less than ten pixels and distant blends as those in which they are separated by ten or more pixels.

To avoid any spurious correlations, blended images are produced by simulating the galaxies individually and then artificially combining them. For simplicity, we only consider blends consisting of two sources. Real images may well contain cases involving more than two sources as well as various artefacts that would make classification more complicated. We leave these issues to be addressed in future work and here focus on testing the applicability of transfer learning to our simple test case.

\subsection{The COSMOS catalogue}
\label{sec:cosmos}

We use the Cosmological Evolution Survey  \citep[COSMOS,][]{scoville:07} catalogue as basis from which to derive our simplistic training data and more realistic testing data. COSMOS was chosen as it provides a representative sample of galaxies in terms of size, ellipticity, luminosity and morphology. COSMOS is a catalogue of Hubble Space Telescope observations of 1.64~$\mathrm{deg}^{2}$ on the the sky with very accurate photometry and morphology. This catalogue contains high resolution images (0.05 arcsec/pixel) with a very small Point Spread Function (PSF) of 0.01 arcsec.

In particular, we use a subset of this catalogue selected and processed for weak lensing purposes \citep{mandelbaum:2012}. This subset of images have a negligible amount of noise and the PSF has been deconvolved. This makes it possible to re-sample the images on a larger pixel scale and to convolve them with a different PSF in order to mimic observations from another instrument. Finally, the galaxies have been fitted by either a Sérsic profile or a bulge + disc profile for which the bulge is represented by a De Vaucouleurs profile \citep{deVaucouleurs:48} and the disc by an exponential profile \citep[see][for details]{mandelbaum:2012}.

The subset of processed COSMOS data has been made available through the Galsim software package \citep{rowe:15}. GalSim is widely used in the astrophysics community to simulate and manipulate galaxy images and was used extensively in several weak lensing challenges, such as GREAT3 \citep{mandelbaum:2014}. The package provides all the tools necessary to work with COSMOS data.

\subsection{Point spread function}
\label{sec:psf}

The impulse response or PSF of an instrument encompasses all of the aberrations induced by the optical system along with other distortions arising from the atmosphere \emph{etc.} The PSF induces blurring in observed images, which artificially increase the size of sources and can cause their light profiles to overlap creating blends. 

We model the optical PSF as a Moffat profile with $\beta=4.765$ \citep{Trujillo:2001} and the PSF ellipticity is drawn from real optical variations of the Canada-France-Hawaii Telescope (CFHT) derived from real CFIS data \citep{guinot:21}. Atmospheric turbulence is modelled by a Kolmogorov profile \citep{tatarski:2016} with random ellipticity drawn from a Gaussian distribution with average $\mu = 0$ and standard deviation $\sigma=0.01$. The final PSF is obtained by convolving the two models and has an average size of 0.65 arcsec. Given the relatively small size of the postage stamps, spatial variations of the PSF are neglected.

This CFIS-like PSF model is simple but sufficiently realistic for the purposes of this work.

\subsection{Parametric model training data}
\label{sec:sims}

In order to keep our training samples as simple and generic as possible, we use a series of parametric models derived from fits to the COSMOS sample described in Sect.~\ref{sec:cosmos}. This constitutes a range of simulated galaxies with different sizes, shapes and ellipticities. The parametric models are then convolved with the CFIS-like PSF described in Sect.~\ref{sec:psf}. Each image corresponds to a $51\times 51$ pixel postage stamp with a convolved galaxy model at the centre. 

For half of the sample a second galaxy model is placed at a random position within the postage stamp to produce blends according to the definition provided in Sect.~\ref{sec:blend_def}.  We pad all of the images with zeros (7 pixels in every direction) to avoid issues when the secondary source is close to the border of the postage stamp.

The complete sample is comprised of 80~000 noiseless, padded postage stamps, half of which are isolated galaxies and the other half are blends.

The final step in producing our simulated training sets is to add Gaussian random noise. In order to test how sensitive blend identification is to noise, we generate 7 different noise standard deviations ($\sigma_{\mathrm{noise}} = 5, 10, 15, 20, 25, 30, 35$). We additionally created 10 realisations of the noise for each value of $\sigma_{\mathrm{noise}}$ in order to test the stability of the training. Each realisation of each noise level is treated as an independent training set. In other words, we train the network 70 times and obtain 70 sets of weights.

\subsection{Realistic testing data}
\label{sec:real}

We generate a sample of realistic CFIS-like images to test our transfer learning approach. To do so, we take the real denoised and deconvolved COSMOS images described in Sect.~\ref{sec:cosmos}, crop them to $51\times 51$ pixel postage stamps and convolve them with the CFIS-like PSF described in Sect.~\ref{sec:psf}. The flux is rescaled in order to reproduce a 300s exposure at the 3.6m CFHT telescope. The images are re-sampled at the resolution of the CFIS survey (0.187 arcsec). Similarly to the training images, blends are created for half of the sample by adding a secondary source at a random position in the postage stamp and then all the postage stamps are zero padded in the same way. Finally, Gaussian noise with $\sigma_{\mathrm{noise}} = 14.5$ was added to the images to replicate the SNR of CFIS data.

The final sample of $80~000$ realistic postage stamps is considerably more complex than the parametric models described in Sect.~\ref{sec:sims} and more closely approximate the conditions expected in real images. For example, the light profiles do not necessarily have central symmetry due to star forming regions or because of complex morphology. Out of the $80~000$ postage stamps $4~838$ correspond to close blends (\emph{i.e.} the sources are less than ten pixels apart) and the remaining $35~162$ correspond to distant blends.

Table~\ref{tab:data} summarises all of the data sets used for training and testing.

\begin{table}[ht]
\caption{Summary of the data used for training and testing.}
\label{tab:data}
\centering
\begin{tabular}{c c c}
 \hline\hline
 Data set & Parametric Training Set & Realistic Testing Set\\
 \hline
 $\mathrm{N}_{\mathrm{isolated}}$ & 40~000    & 40~000\\
 $\mathrm{N}_{\mathrm{blended}}$ &   40~000  & 40~000\\
 $\sigma_{\mathrm{noise}}$ & 5, 10, 15, 20, 25, 30, 35 & $14.5$\\
 $\mathrm{N}_{\mathrm{\sigma\ real}}$ &   10  & 1\\
 \hline
\end{tabular}
\end{table}

\noindent $\mathrm{N}_{\mathrm{isolated}}$ is the number postage stamps containing isolated sources, $\mathrm{N}_{\mathrm{blended}}$ is the number of postage stamps containing blended sources, $\sigma_{\mathrm{noise}}$ is the amount of noise added to the postage stamps and $\mathrm{N}_{\mathrm{\sigma\ real}}$ is the number of realisations for each noise level.


\section{Deep transfer learning framework for blend identification}
\label{sec:framework}

The objective of transfer learning is to learn a set of weights from a given training set and then transfer these to a related but independent problem. This is interesting from the perspective of blend identification given that we do not have a large sample of labelled data and training with a small sample of known blends may significantly bias the learned weights. Additionally, the use of pre-trained weights considerably reduces the time required to train a network, thereby making the application of a deep learning approach to galaxy survey data more feasible.

 For the purposes of this work we test the applicability of transfer learning to the problem of blend identification in a two-step process.  In the first step, CNN weights learned from natural images are used to extract generic features from the galaxy images. The fully connected layers are then trained on galaxy image features derived from simple parametric models. The objective being to capture the general properties of blended images and not specific features of the individual galaxies. In the second step, the weights learned from simple parametric models are applied to more realistic CFIS-like images to test the classification accuracy.

The architecture we choose to implement our deep transfer problem was that of VGG-16. For simplicity we refer to our specific VGG-16 set up for the problem of blend identification as \textsc{BlendHunter}.

\subsection{The VGG-16 network}

\begin{figure}
    \centering
    \includegraphics[width=0.5\textwidth]{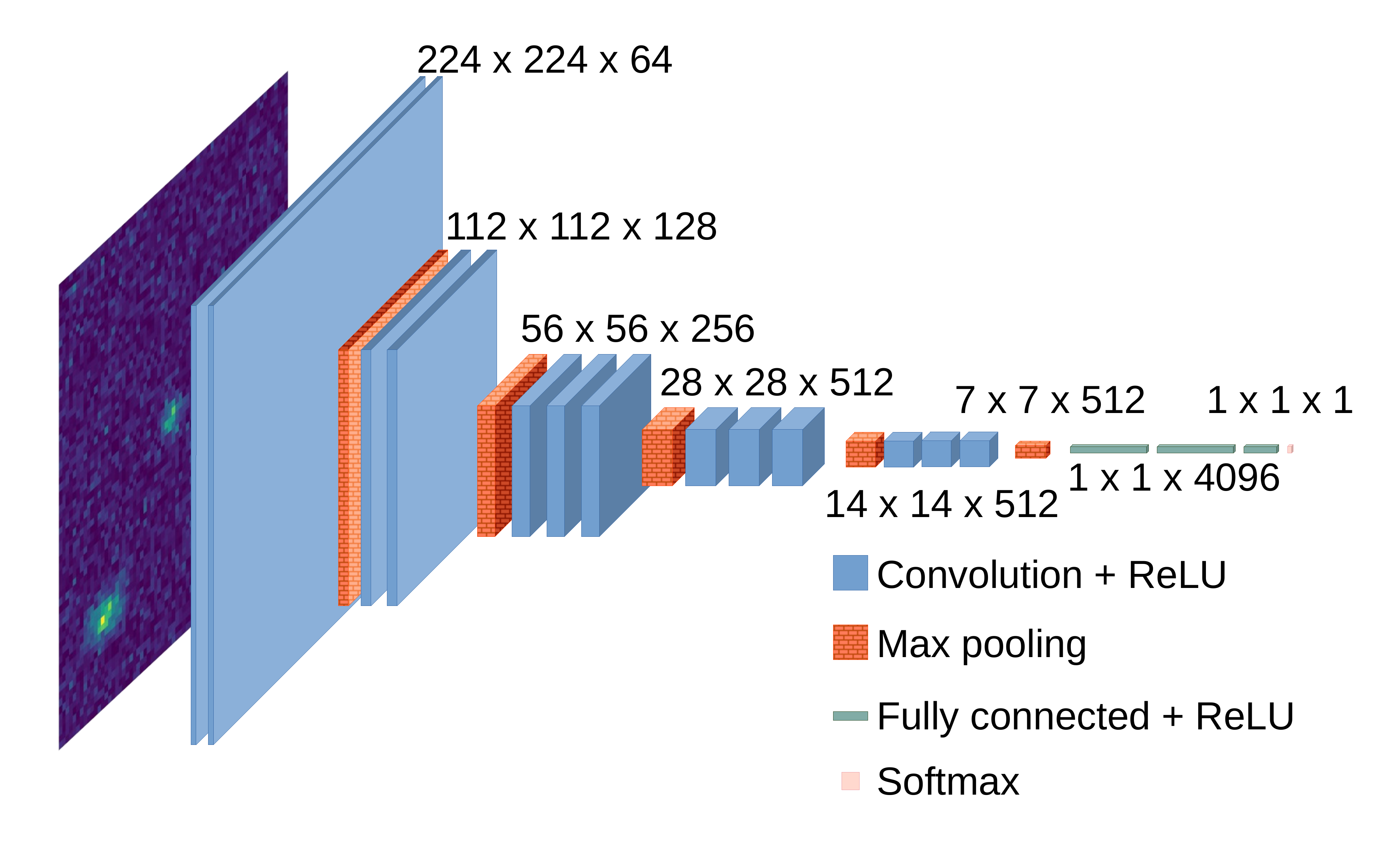}
    \caption{Visual representation of the VGG-16 network. Convolutional layers with ReLU activation are shown in solid blue, max pooling layers are shown in brick-pattern red, fully connected layers are shown as green bars and the output softmax layer is shown as the last box in pink.}
    \label{fig:vgg16_arch}
\end{figure}

VGG-16 is a deep convolutional network with 16 weight layers developed by the Visual Geometry Group (VGG) at the University of Oxford \citep{vgg16}. The network was ranked first in the ImageNet Large-Scale Visual Recognition Challenge (ILSVRC) in 2014 \citep{russakovsky:15}. The main feature of this architecture was the increased depth of the network compared to the state of the art at the time. In VGG-16, three-channel images (RGB) are passed through 5 blocks of convolutional layers, where each block is composed of increasing numbers of $3\times 3$ filters. The stride (\textit{i.e.} the amount by which the filter is shifted) is fixed to 1, while the convolutional layer inputs are padded such that the spatial resolution is preserved after convolution (\textit{i.e.} the padding is 1 pixel for $3\times 3$ filters). The blocks are separated by max-pooling (\textit{i.e.} down-sampling) layers. Max-pooling is performed over $2\times 2$ windows with stride 2. The 5 blocks of convolutional layers are followed by three fully-connected layers. The final layer is a soft-max layer that outputs class probabilities. The full network architecture used is shown in Fig.~\ref{fig:vgg16_arch}.

The VGG-16 network was chosen for purposes of the work presented here for several reasons. Firstly, the network can be implemented with weights pre-trained on the ImageNet database \citep{deng:09} in order to save computation time and resources. The diversity of this data set has allowed the network to learn a variety of generic image features, which are applicable to most image classification tasks. Finally, VGG-16 has already been applied to a variety of astrophysical applications, including galaxy morphology classification \citep{wu:19, zhu:19}, glitch classification in gravitational wave data \citep{george:18} and coronagraph image classification \citep{shan:20}.

The VGG-16 network was implemented in Python via the TensorFlow-Keras neural network API \citep{chollet:15, tensorflow:2015}, which includes ImageNet pre-trained weights.

\subsection{Training and validation}
\label{sec:training}

We trained BlendHunter in two phases using the 70 sets of 80~000 simulated images described in Sect.~\ref{sec:sims}. We divided each data set in the follow way: 36~000 images were used for training, 36~000 for validation and the remaining 8~000 for testing.

\subsubsection{Convolutional layers}

In the first phase, the we simply initialise the convolutional layers (\emph{i.e.} the first 13 weight layers) with the pre-trained ImageNet weights. As mentioned in the previous section, the purpose of using pre-trained weights is to save processing time. Therefore, this part of the network is not actually trained with our data and can be seen as a simple feature extractor. 

As ImageNet weights are used, the VGG-16 network expects RGB images as inputs. Therefore, we rescaled the simulated images to the range $i \in \mathbb{Z} : 0 \leq i \leq 255$ and the monochromatic pixel values were repeated across the 3 RGB channels. The final images are saved as Portable Network Graphics (PNG) files which are fed into the network.  

Given the relatively small amount of training data, we additionally implemented data augmentation in order to obtain more diversified features from the convolutional layers of the network. Augmenting our simulated images simply means creating additional images with minor changes such as flips, translations or rotations. Specifically, we used the Keras image pre-processing modules to include a shear range and zoom range of $0.2$, as well as horizontal flipping. Note that data augmentation was not used on the test images.

\begin{figure*}
    \centering
    \includegraphics[width=0.9\textwidth]{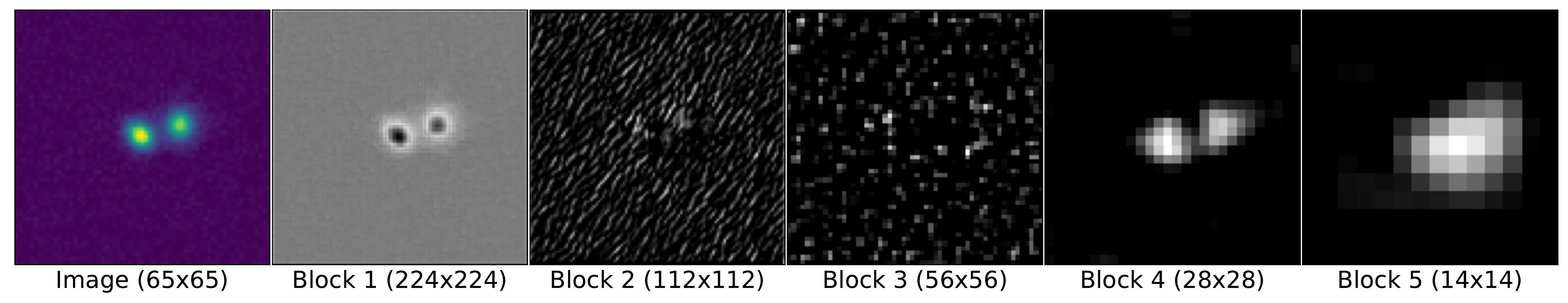}
    \caption{Examples of features extracted from the VGG-16 convolutional layers pre-trained with ImageNet weights. The leftmost panel shows an input postage stamp containing blended sources. The following panels show example features extracted from various convolution blocks.}
    \label{fig:block1_features}
\end{figure*}

Fig.~\ref{fig:block1_features} shows some examples of features extracted from on of the simulated images with the convolutional layers. The earlier convolutional blocks provide more general features while the later blocks provide more specific features.

\subsubsection{Fully connected layers}
\label{sec:connected}

In the second phase, we train the fully connected layers (\emph{i.e.} the last 3 weight layers) on top of the previously stored features. These layers consist of a layer with linear transformation, followed by a layer with dropout and finally a layer with Rectified Linear Unit (ReLu) activation. We chose this non-linear activation function, widely used with deep neural networks, because it allows for faster convergence. We used dropout alongside data augmentation to avoid over-fitting \citep{dropoutHinton}. This is a way of cancelling some activations to prevent the network from learning random correlations in the images. We employed a dropout rate of $0.1$ for our experiments. For the training, we proceeded with batches of 250 images at a time over 500 epochs. An epoch corresponds to passing the entire training sample through the network and a batch size of 250 proved to be the most computationally efficient for the gradient descent. 

We use the binary cross entropy loss function in Eq. \ref{eq:bce}

\begin{ceqn}
\begin{align}
    \mathrm{BCE} = - \frac{1}{N} \sum_{i = 1}^{N} y_i\cdot \log(p(y_i)) + (1 - y_i)\cdot \log(1 - p(y_i))
\label{eq:bce}
\end{align}
\end{ceqn}

\noindent where $y_i$ is the true label (1 for blended, 0 otherwise) for the $i^{th}$ image and $p(y_i)$ is the probability of the image being blended or not returned by the network. We selected this loss function to optimise as it is typically used in this type of binary classification problem.

The Adam algorithm \citep{kingma:14} was used to minimise our loss function. In contrast to a classic Stochastic Gradient Descent (SGD) algorithm, it computes individual learning rates for each weight and is quite efficient in the optimisation of deep neural networks. In order to adapt the learning rate, Adam requires the estimations of the first (the mean) and second (the uncentred variance) moments of the gradient using exponentially moving averages (Eqs.~\ref{eq:mean} and \ref{eq:var}, respectively), with $g_t$ the objective function gradient at time step $t$, and $\beta_{1}$ and $\beta_{2}$ the exponential decay rates for the moment estimates.

\begin{ceqn}
\begin{align}
    m_t = \beta_{1} \cdot m_{t-1} + (1- \beta_{1}) \cdot g_t 
\label{eq:mean}
\end{align}
\end{ceqn}

\begin{ceqn}
\begin{align}
    v_t = \beta_{2} \cdot v_{t-1} + (1- \beta_{2}) \cdot g_t^{2}
\label{eq:var}
\end{align}
\end{ceqn}

\noindent It takes advantage of both RMSprop \citep{tieleman:2012} and AdaGrad \citep{duchi} methods. In addition to being robust and less time-consuming, Adam can thus be applied to a wider selection of optimisation problems. It also requires almost no tuning of its parameters. We set $\beta_1$ and $\beta_2$ to their default values, respectively 0.9 and 0.999. Finally, the weights are updated according to Eq.~\ref{eq:update_w}, where $w_t$ are the fully-connected network weights at time step $t$, $\eta$ is the step size, $\hat{m_t}$ and $\hat{v_t}$ are the bias corrected estimators of the first and second moments, and $\epsilon$ a value set to $10^{-8}$ to prevent division by 0.

\begin{ceqn}
\begin{align}
    w_t = w_{t-1} - \eta\cdot\frac{\hat{m_t}}{\sqrt{\hat{v_t}} + \epsilon}
\label{eq:update_w}
\end{align}
\end{ceqn}

At the end of each epoch, both training and validation losses were computed, and the weights were updated every time the validation loss decreased. Training was stopped when the validation loss had not decreased after 10 epochs. The network converged after around 70 epochs on average.

We began training with a learning rate of $1.10^{-3}$. Since choosing the right learning rate can be challenging, we decided to reduce the learning rate by a factor of 0.5 every time the validation loss did not decrease after 5 epochs. A small learning rate would make it possible to avoid big jumps in gradient descent. Otherwise, in this case, it could fail to converge and settle around a local minimum. No weight decay was implemented in this phase.

Tuning deep neural networks hyperparameters with a considerable amount of parameters to learn can prove to be very time-consuming. This is why we focused on the hyperparameters that would have the most impact on the results. Several tests were made such as changes to the regularisation, weight initialisation, dropout rate, learning rate and optimiser. However, no significant improvement in accuracy was observed. Switching to the SGD optimiser or increasing the dropout rate led to worse performance overall.

The network takes approximately 630s to train on a sample of 80~000 images using a standard Intel(R) Core(TM) i7-6900K CPU (3.20GHz, 32GB).


\section{Results}
\label{sec:comparison}

\subsection{SExtractor benchmark}

We compare the performance of \textsc{BlendHunter} with \textsc{SExtractor}, as this is the most widely used tool in the community for identifying and handling blended sources in astronomical images. The objective being to test the reliability of our approach versus the state of the art. Specifically, We make use of the \textsc{SExtractor} Python wrapper \textsc{SEP} \citep{barbary:2016} for our tests. Note that \textsc{SEP} does many things beyond blend identification and many of these steps can not easily be isolated, however we tried to make the comparison as fair and consistent as possible.

\textsc{SEP} implements a multi-thresholding technique to decompose detected sources into sub-sources (when possible). This method takes two input parameters: the number of bins to decompose the light profile and the minimum contrast value between the main peak and a given sub-peak. The contrast is evaluated based on the flux of each peak (see Fig.~2 in \cite{bertin:96}). The set of parameters we used for \textsc{SEP} is shown in Table~\ref{table:sex_param} (all the other parameters are kept to their default values). The value of $0.005$ for \texttt{DEBLEND\_MINCONT} (the minimum contrast parameter for deblending) may be considered a little high compared to that commonly used in the literature, but here we chose to favour reliable identification over increasing the number of blends found at the cost of also increasing the number of spurious detections.

\begin{table}
\caption{\textsc{SEP} parameter settings.}
\label{table:sex_param}
\centering
\begin{tabular}{p{3cm} p{4.3cm}}
	\hline\hline
	Parameter & Value \\
	\hline
	\texttt{THRESH\_TYPE} & RELATIVE \\
	\texttt{DETECT\_THRESH} & 1.5 \\
	\texttt{DETECT\_MINAREA} & 5 \\
	\texttt{FILTER} & Y \\
	\texttt{FILTER\_NAME} & kernel\_3x3.conv (default) \\
	\texttt{DEBLEND\_NTHRESH} & 32 \\
	\texttt{DEBLEND\_MINCONT} & 0.005 \\
	\hline
\end{tabular}
\end{table}
 
Given our loose definition of blended sources (see Sect.~\ref{sec:blend_def}), we chose not to rely exclusively on the deblending flags provided by \textsc{SEP}. Instead, we also check that the sources are found at the right positions (within a two pixel radius) to make sure we do not extract noise features. Additionally, since some of the sources in the postage stamp do not technically overlap, when \textsc{SEP} correctly identifies the number of sources (\emph{i.e.} 1 or 2) we take this as a correctly labelled postage stamp.

The process for labelling postage stamps as either isolated or blended sources using \textsc{SEP} can be summarised as follows:

\bigskip

\noindent If a single source is detected:
\begin{itemize}
    \item If the source is flagged as a blend and is at the expected position, the image is  labelled as a blend.
    \item Otherwise, the image is labelled as an isolated source.
\end{itemize}

\noindent If two sources are detected:
\begin{itemize}
    \item If both sources are detected at the right positions, the image is labelled as a blend (this stands even if one of the two sources is itself detected as a blend).
    \item If the sources are incorrectly identified, the image is labelled as an isolated source.
\end{itemize}

\noindent If more than two sources are detected:
\begin{itemize}
    \item If at least the right number of sources are detected at their expected positions, the image is labelled as a blend.
    \item If the number of detected sources is larger than the number of true sources, the image is flagged as a \textsc{SEP} failure and is not included in the analysis. Note that this occurred for less than $1\%$ of the images. 
\end{itemize}

\begin{figure*}[ht]
    \centering
    \includegraphics[width=0.95\textwidth]{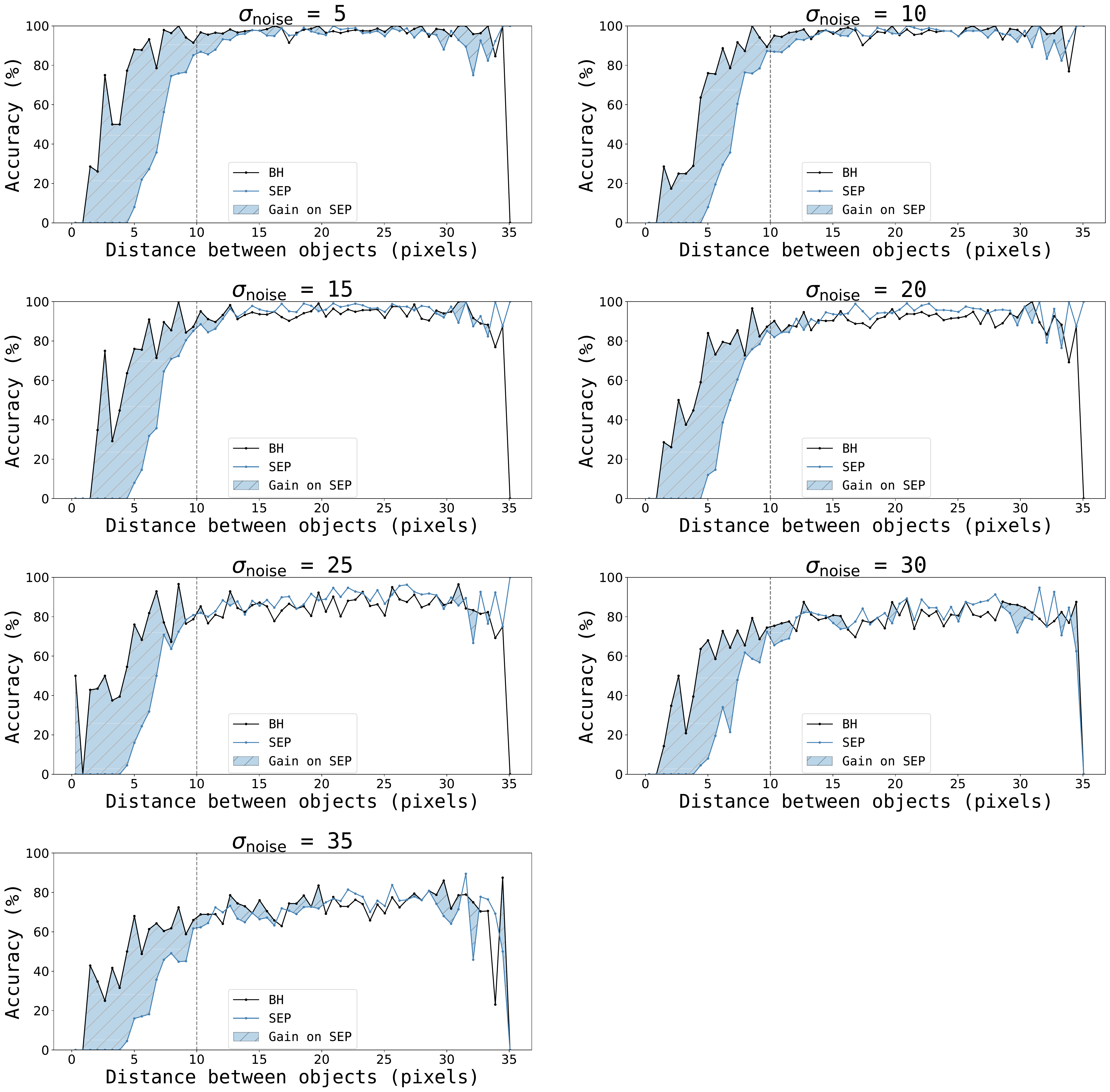}
    \caption{Classification accuracy of \textsc{BlendHunter} (black solid line) versus \textsc{SEP} (blue dashed line) as a function of separation between sources. Results are from the blended samples of the COSMOS parametric model testing sets. Each panel shows one realisation of a given noise standard deviation, $\sigma_{\mathrm{noise}}$. The blue shaded area shows the gain in accuracy of \textsc{BlendHunter} with respect to \textsc{SEP}.}
    \label{fig:distances}
\end{figure*}

\subsection{Results from parametric models}

In the following, we define classification accuracy as the percentage of postage stamps correctly labelled as either isolated or blended sources. Here we take the weights obtained via the training procedure described in \ref{sec:training} and apply them to the test sample of $8~000$ parametric model postage stamps (see Sect.~\ref{sec:sims}) to obtain classification labels.

Fig.~\ref{fig:distances} shows the classification accuracy of \textsc{BlendHunter} (black solid line) versus \textsc{SEP} (blue dashed line) as a function of separation between sources. Results are from the blended samples of the COSMOS parametric model testing sets (\emph{i.e.} sets of $4~000$ postage stamps). Each panel shows one realisation of a given noise standard deviation, $\sigma_{\mathrm{noise}}$. The blue shaded area shows the gain in accuracy of \textsc{BlendHunter} with respect to \textsc{SEP}. For close blends (\emph{i.e.} where the two sources are less than 10 pixels apart), \textsc{BlendHunter} significantly outperforms \textsc{SEP} showing gains as high as $\sim80\%$ in classification accuracy. For distant blends, the two techniques appear to be consistent to within a few percent and correctly label the majority of the postage stamps.

\begin{figure}[ht]
    \centering
    \includegraphics[width=0.49\textwidth]{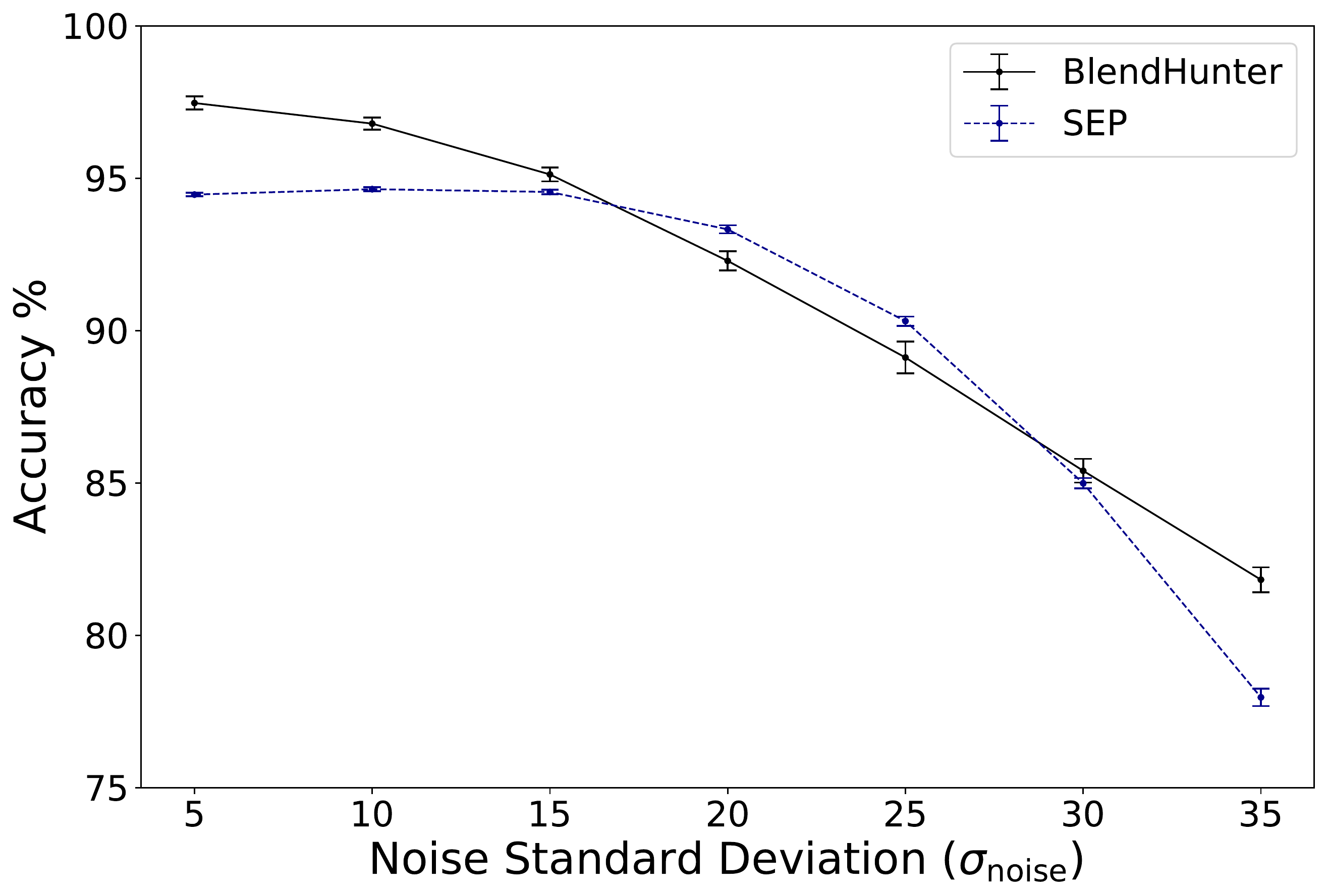}
    \caption{Overall classification accuracy of \textsc{BlendHunter} (black solid line) versus \textsc{SEP} (blue dashed line) with respect to $\sigma_{\mathrm{noise}}$ on the COSMOS parametric model testing set. The points are taken from the average accuracy from ten realisations of each noise level and the error bars from the standard deviation.}
    \label{fig:overall_acc}
\end{figure}

In Fig.~\ref{fig:overall_acc} we display the overall accuracy as function of noise standard deviation for \textsc{BlendHunter} and \textsc{SEP} on the full parametric model testing set (\emph{i.e.} sets of $8~000$ postage stamps). The points are taken from the average classification accuracy from the ten noise realisations and the error bars from the standard deviation. As expected, both approaches perform almost perfectly for low noise levels and drop off as the noise increases. Overall the two approaches appear fairly consistent, however \textsc{BlendHunter} drops off less rapidly for higher noise levels indicating it may be slightly more robust in higher noise regimes. 

\subsection{Results from realistic CFIS-like images}
\begin{figure}[ht]
    \centering
    \includegraphics[width=0.49 \textwidth]{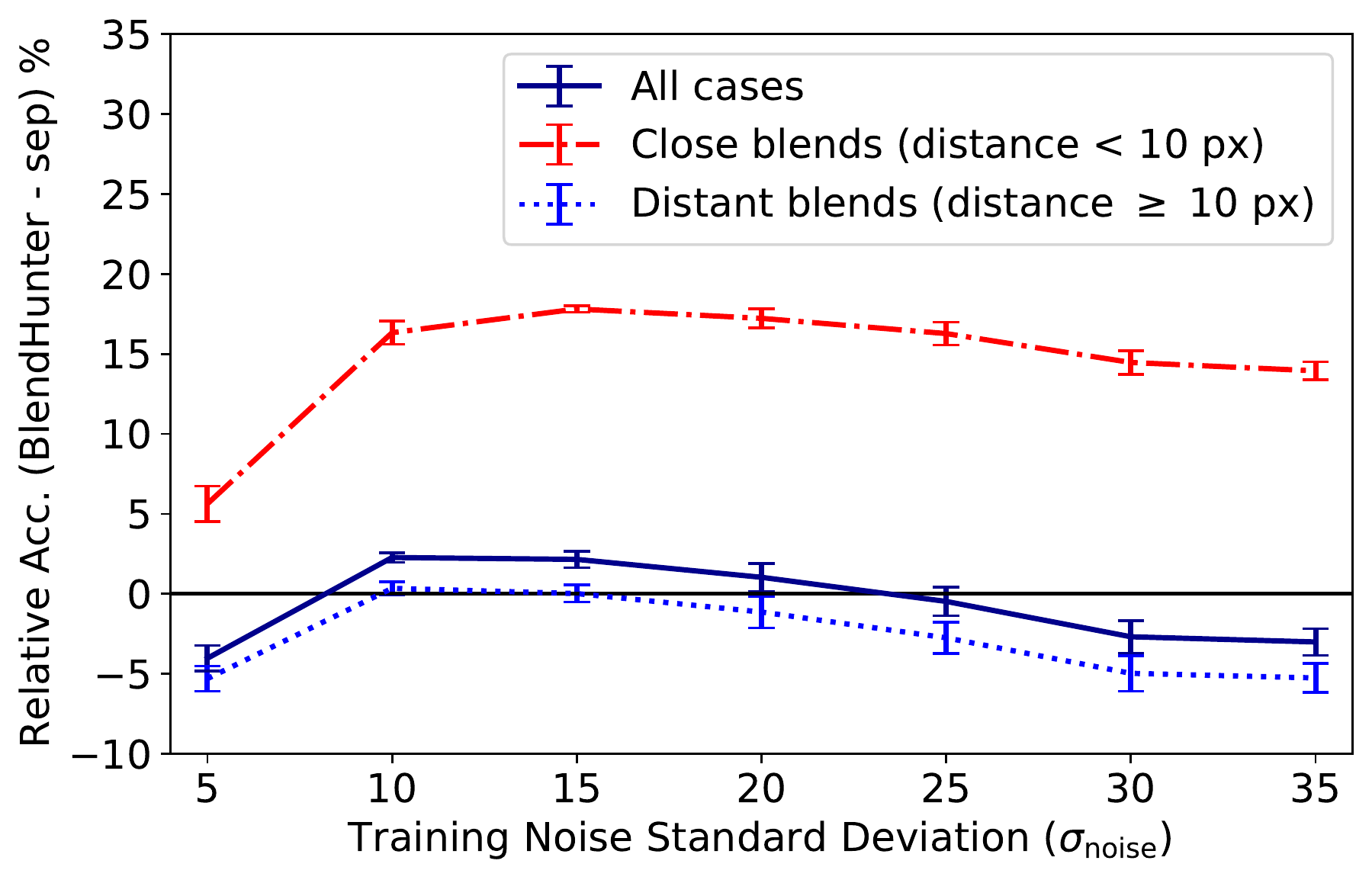}
    \caption{Relative classification accuracy for \textsc{BlendHunter} with respect to \textsc{SEP} on the realistic CFIS-like postage stamps, with $\sigma_{\mathrm{noise}} = 14.5$. The points are taken from the average relative classification accuracy from the ten training noise realisations and the error bars from the standard deviation. For close blends (red dot-dashed line), \textsc{BlendHunter} outperforms \textsc{SEP} for any level of noise in the training data.}
    \label{fig:cfis_like}
\end{figure}

Here we test the weights obtained from the parametric models with various noise levels and apply them to the test sample of $80~000$ realistic CFIS-like postage stamps (see Sect.~\ref{sec:real}) to obtain classification labels.

Fig.~\ref{fig:cfis_like} shows the relative classification accuracy as function of the training noise standard deviation (\emph{i.e.} the noise level of the sample used for training) for \textsc{BlendHunter} with respect to \textsc{SEP} on the realistic CFIS-like postage stamps. Points show the average relative classification accuracy corresponding to the ten noise realisations for each training noise level and the error bars are the standard deviation. \textsc{SEP} is simply run once on the full data set. For the whole set of images (\emph{i.e.} all $80~000$ postage stamps), we can see that \textsc{BlendHunter} is able to slightly outperform \textsc{SEP} (only by a few percent) around $\sigma_{\mathrm{noise}} = 14.5$, which is the true noise standard deviation of the CFIS-like sample. The classification accuracy then degrades if the network is trained on noise levels that differ significantly. In the subset of close blends ($4~838$ postage stamps), \textsc{BlendHunter} significantly outperforms SEP regardless of the training conditions. This is consistent with the results from the parametric model data.  Finally, for distant blends ($35~162$ postage stamps), \textsc{BlendHunter} matches the performance of SEP when trained on the same noise level as the target images and under-performs by a maximum of $5\%$ when trained on a noise level that differs. The overall accuracy of SEP in each case is: $91\%$ over the whole data set, $69\%$ for close blends, and $94\%$ for distant blends. 

These results indicate that \textsc{BlendHunter} is not overly sensitive to the training without perfect knowledge of the noise in the target sample. This is a strong indication that the transfer learning has been successful and promising for the prospect of applying this approach to real survey data. Additionally, the performance of \textsc{BlendHunter} is noticeably more robust for close blends, where galaxy profiles significantly overlap. 

It should be stressed that these results are based on the relatively simple blend definition provided in section~\ref{sec:blend_def}, which does not take into account all of the complexities that could be encountered for real extended sources.

\begin{figure*}[ht]
    \centering
    \includegraphics[width=0.3 \textwidth]{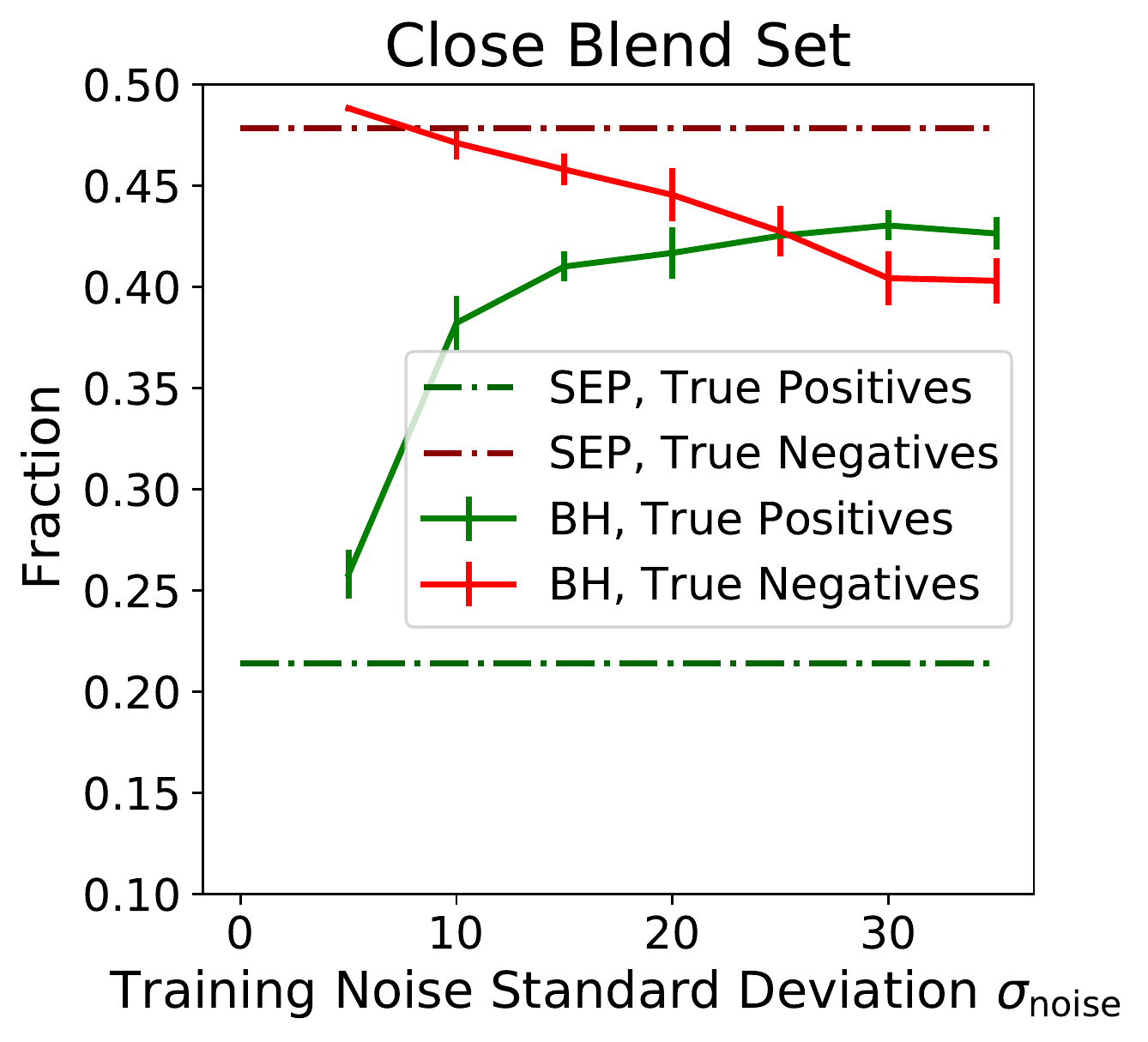}  \includegraphics[width=0.3 \textwidth]{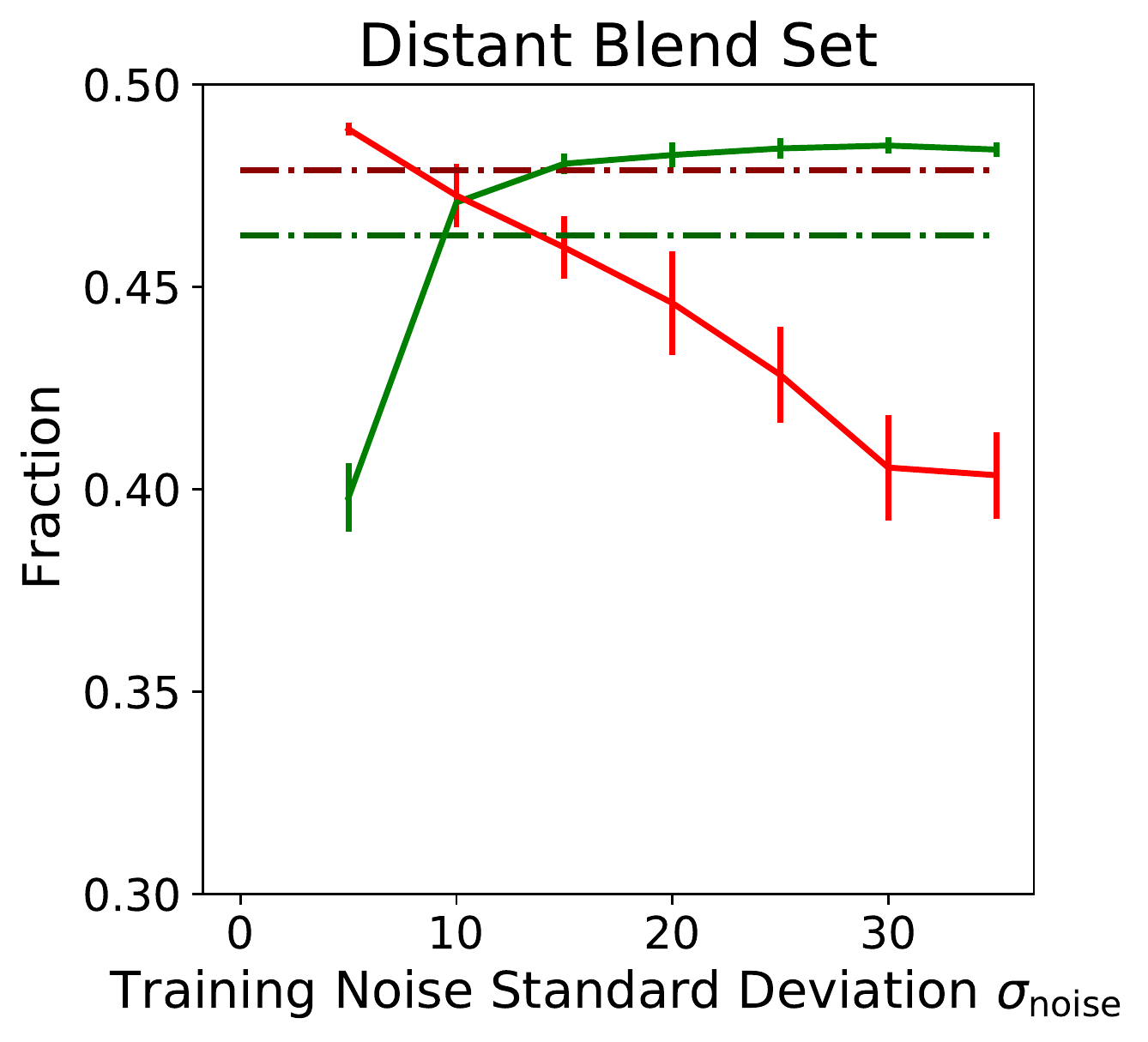}    \includegraphics[width=0.3 \textwidth]{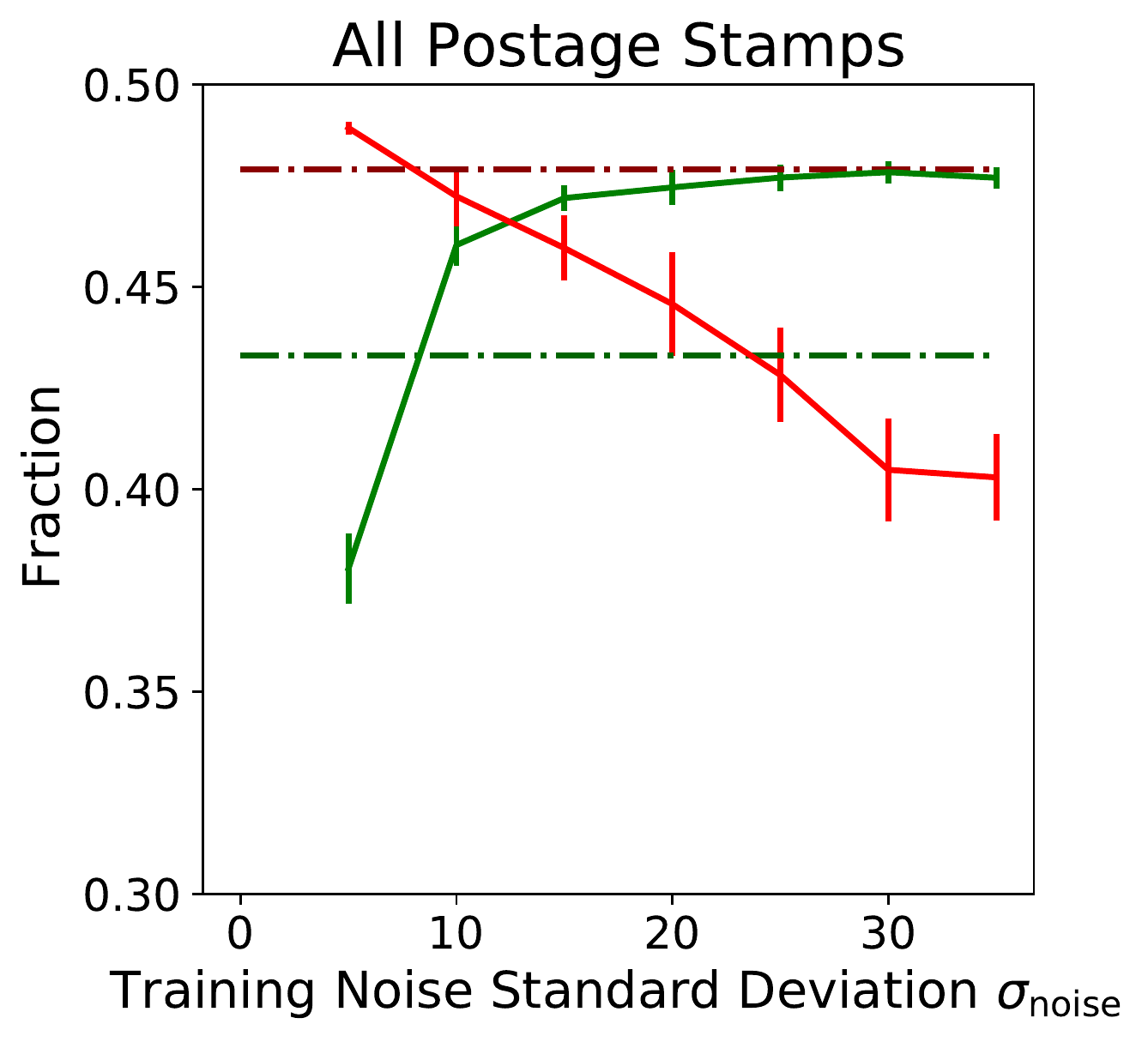}
    \caption{Classification confusion matrices of \textsc{BlendHunter} (solid lines) compared to \textsc{SEP} (dot-dashed lines) for the realistic CFIS-like test data. Each panel shows the fraction of blends correctly classified as blends (\emph{i.e.} true positives, green lines) or isolated sources correctly classified as isolated sources (\emph{i.e.} true negatives, red lines) as a function of the the training noise standard deviation. The left panel shows the sample of close blends along with an equivalent number of isolated sources. The middle panel shows the sample of distant blends along with the remaining isolated sources. The right panel shows the results for the full set of postage stamps.}
    \label{fig:confusion}
\end{figure*}

We additionally examined the confusion matrices for the predictions from the realistic CFIS-like test set. Fig.~\ref{fig:confusion} shows the fraction of blends correctly classified as blends (\emph{i.e.} true positives) or isolated sources correctly classified as isolated sources (\emph{i.e.} true negatives) as a function of the the training noise standard deviation. The first panel shows the performance of \textsc{BlendHunter} compared to \textsc{SEP} for the sample of close blends (green lines) along with an equivalent number (4~838) of isolated sources (red lines). The second panel shows the equivalent plot for distant blends with the remaining sample of isolated sources. The third panel shows the results for all 80~000 postage stamps.

The results show that \textsc{BlendHunter} is significantly better at identifying true blends than \textsc{SEP} regardless of the noise level used for training. This is particularly noticeable for the sample of close blends. \textsc{BlendHunter}, however, also produces more incorrect labels for isolated sources and this worsens as the training noise level increases beyond the that of the target sample.

\section{Conclusions}
\label{sec:conclusions}

We have presented a proof-of-concept deep transfer learning approach for the automated and robust identification of blended sources in galaxy survey data called \textsc{BlendHunter}. This technique uses convolution layers pre-trained on natural images from ImageNet, thus significantly reducing the time required for training. The fully connected layers are then trained on simple parametric models derived from COSMOS data for various levels and realisations of Gaussian random noise.

Comparison with the community standard \textsc{SEP} (a Python implementation of \textsc{SExtractor}) on 70 sets of $8~000$ parametric models demonstrate that \textsc{BlendHunter} is significantly more sensitive to blends when the images are very noisy or when galaxy pairs are very close ($<10$ pixels). In the low noise, distant blend regime, \textsc{BlendHunter} appears to be consistent with \textsc{SEP}.

We additionally produced a sample of $80~000$ more realistic CFIS-like images derived from real COSMOS data. Results demonstrate that the \textsc{BlendHunter} weights, learned on the parametric models, can successfully be transferred to this more complex data set and achieve $>90\%$ classification accuracy, outperforming \textsc{SEP} by a few percent on the full sample when the appropriate weights are used. The results also indicate that \textsc{BlendHunter} is capable of achieving results roughly consistent with \textsc{SEP} even when weights are used that have been trained with a significantly different noise standard deviation.

\textsc{BlendHunter} notably outperforms \textsc{SEP} by $5-15\%$ for blends in which the galaxies are separated by less than ten pixels. This is a interesting result as these are precisely the cases that generate biases in galaxy shape measurements \citep{MacCrann:2020}. 

Overall the results are promising and indicate that it may be possible to adapt this approach to more accurately identify blended sources in real survey data. The next steps for moving in this direction would entail generating more realistic testing data that contain some artefacts and images with more than two sources. Further work is also required to reduce the number of false negatives, \emph{i.e.} incorrect labels for isolated sources. Finally, additional tests should be performed to determine if and to what degree the use of pre-trained weights in the CNN layers can help prevent over-fitting the network to the training sets. We leave the investigation and implementation of these steps for future work.


\begin{acknowledgements}
    The authors wish to acknowledge the COSMIC project funded by the CEA DRF-Impulsion call in 2016, the CrossDisciplinary Program on Numerical Simulation (SILICOSMIC project in 2018) of CEA, the French Alternative Energies and Atomic Energy Commission. The Euclid Collaboration, the European Space Agency and the support of the Centre National d’Etudes Spatiales. This work was also supported by the ANR AstroDeep project - grant 19-CE23-0024-01. This work has made use of the CANDIDE Cluster at the Institut d'Astrophysique de Paris and made possible by grants from the PNCG and the DIM-ACAV. This work is based on data obtained as part of the Canada-France Imaging Survey, a CFHT large program of the National Research Council of Canada and the French Centre National de la Recherche Scientifique. Based on observations obtained with MegaPrime/MegaCam, a joint project of CFHT and CEA Saclay, at the Canada-France-Hawaii Telescope (CFHT) which is operated by the National Research Council (NRC) of Canada, the Institut National des Science de l'Univers (INSU) of the Centre National de la Recherche Scientifique (CNRS) of France, and the University of Hawaii. This research used the facilities of the Canadian Astronomy Data Centre operated by the National Research Council of Canada with the support of the Canadian Space Agency. AZV would like to thank LSC/PMR/EP at University of São Paulo for providing additional computing power. The authors also wish to thank Xinyu Wang and Alexandre Bruckert for initial efforts that were expanded upon in this work.
\end{acknowledgements}


\bibliographystyle{aa}
\bibliography{blendhunter}


\end{document}